%% file: main.tex
\documentclass{article}
\usepackage[T1]{fontenc}
\usepackage[utf8]{inputenc}
\usepackage{ismir}

\usepackage{amsmath,amssymb,cite,url}
\usepackage[hidelinks,implicit=false,bookmarks=false]{hyperref}
\usepackage{graphicx}
\usepackage[table]{xcolor}
\usepackage{booktabs,multirow}

\definecolor{rowhl}{HTML}{E8F0FE}    
\definecolor{sechl}{HTML}{F0F0F0}    
\definecolor{best}{HTML}{B30000}      
\definecolor{second}{HTML}{0B3D91}   
\newcommand{\first}[1]{\textcolor{best}{\textbf{#1}}}
\newcommand{\second}[1]{\textcolor{second}{\underline{#1}}}

\usepackage{caption}
\title{CODA: \underline{C}ascaded \underline{O}nline \underline{D}iscontinuity-Aware
\underline{A}lignment for Real-Time Image-Based Score Following}

\multauthor
  {Yining Yang$^{1*}$ \hspace{0.5cm} Ruogu Chen$^{2*}$ \hspace{0.5cm} Jie Han$^2$}
  {$^1$ Don Wright Faculty of Music, Western University, Canada\\
   $^2$ Department of Electrical and Computer Engineering, University of Alberta, Canada\\
   {\tt\small yyang524@uwo.ca, ruogu@ualberta.ca, jhan8@ualberta.ca}\\[2pt]
   \small $^*$Equal contribution}

\def\authorname{Y. Yang, R. Chen, and J. Han}

\begin{document}

\maketitle
\thispagestyle{empty}

\begin{abstract}
Real-time score following from sheet images remains challenging because the model must process streaming audio while resolving highly repetitive visual patterns under strict latency constraints. Recent image-based methods have attempted to use multi-resolution prediction by simultaneously predicting the positions of the active system, bar, and note. However, their predictions across these different levels of notation are independent, which makes the predictions unstable and introduces unnecessary extra search space for bar- and note-level predictions. Most existing methods also lack mechanisms to recover from score discontinuities, such as repeats, da capo (D.C.), or coda jumps. This paper proposes CODA, to the best of our knowledge, the first real-time score following system that addresses both gaps. CODA explicitly exploits the cascaded structure of music scores: it first selects the active system, then the active bar within it, and finally the active note within the selected bar. This enforces prediction consistency across resolutions. A silence-driven break mode enables recovery from arbitrary score discontinuities without requiring knowledge of the repeat structure. Evaluated on the Multimodal Sheet Music Dataset (MSMD) piano benchmarks, CODA achieves state-of-the-art tracking accuracy and discontinuity-recovery performance under real-time throughput. Code is available at \url{https://github.com/ValleyC/CODA}.
\end{abstract}

\input{sections/introduction}
\input{sections/related_work}
\input{sections/method}
\input{sections/experiments_and_results}
\input{sections/conclusion}
\input{sections/acknowledgments}

\bibliography{references}

\end{document}

%% file: sections/introduction.tex
\section{Introduction}\label{sec:introduction}

Real-time score following is the process of aligning a live performance with the corresponding music score as the music proceeds in real time. It is fundamental for downstream applications, including automatic accompaniment, automatic page-turning, synchronized score display, and many other interactive music systems \cite{orio2003score,dannenberg2006music}. Classical approaches use online Dynamic Time Warping (DTW) or probabilistic state models to achieve real-time alignment \cite{dixon2005line,cont2010coupled,nakamura2016real}. However, these methods require a symbolic score such as MIDI or MusicXML, which is not always available.

Image-based score following has emerged over the past decade by working directly on sheet images. Early work classified positions into coarse staff-level buckets on small local image crops \cite{dorfer2016towards}, limiting both spatial resolution and the field of view. Reinforcement-learning methods widened the view to score strips but still operated on narrow windows, making re-engagement difficult when the tracker drifted off track \cite{dorfer2018rl,henkel2019score}. Full-page segmentation addressed the field-of-view limitation by predicting a pixel-level heatmap over the entire page \cite{henkel2020learning}, but only localized at the note level without explicit system or bar tracking. The most recent of these image-based methods proposed multi-resolution detection with You Only Look Once (YOLO)-based object detection frameworks \cite{henkel2021real}. This work predicts the positions of systems, bars, and notes simultaneously using three independent output heads. However, there are no constraints or cascaded structures among these predictions. Nothing enforces that the predicted note lies within the predicted bar, or that the predicted bar lies within the predicted system. The model implicitly learns structural consistency, and disagreements among the three prediction levels are often observed.

This independence causes another problem. Because each head independently predicts over the full page, the bar and note heads must search all candidates on the page, increasing the likelihood of errors and instabilities. A cascaded structure would narrow the search at each stage: once the system is selected, only bars within that system are candidates; once a bar is selected, the candidate note positions are confined to that bar. More fundamentally, this method frames score tracking as an object detection problem. It detects from scratch at every audio frame, predicting bounding boxes as if the system and bar locations were unknown. However, in music performances, the score is static, and all candidate positions are known in advance. Detection is therefore unnecessary. The real task is to select which of the known candidates is currently being played. This formulation is feasible whenever score layout information is available. 

Additionally, existing image-based methods lack mechanisms to recover from score discontinuities such as repeats, da capo (D.C.), or coda jumps, even though such events are common in practice and have long been addressed in symbolic score following \cite{nakamura2016real,morsi2022bottlenecks,park2025matchmaker}. In our analysis of the MSMD test set, 66 out of 94 pieces contain written repeat structures, producing 131 jumps in standard performance order. Despite this prevalence, no existing real-time image-based score following method includes an explicit mechanism for handling such events during online tracking.

This paper proposes CODA, a cascaded online alignment framework that addresses these gaps. The key observation is that on a fixed score page, all system and bar locations are already known. The tracking problem is therefore reformulated as a selection task among known candidates. CODA follows the cascaded structure of music scores directly: it first selects the active system among all systems, then the active bar within that selected system, and finally the note position inside the selected bar. This formulation naturally enforces geometric consistency and narrows the search space at each stage. To handle score discontinuities, CODA uses silence as an observable cue for arbitrary position changes, enabling a single recovery mechanism for repeats, D.C., coda jumps, and performer errors without requiring knowledge of the score's repeat structure. The major contributions are:
\begin{itemize}
\item A cascaded selection-and-regression formulation that enforces geometric consistency across system, bar, and note predictions while narrowing the search space at each stage.
\item A causal streaming architecture combining Mamba audio encoding, FiLM conditioning, cross-attention, and beam search with learned temporal priors.
\item A silence-driven jump recovery mechanism for arbitrary score discontinuities, including repeats, D.C., and coda jumps, without requiring prior knowledge of the score's repeat structure.
\item A repeat-aware jump test benchmark with both annotated and random music discontinuities for all 94 pieces in the MSMD test set, providing the first standardized evaluation protocol for discontinuity handling in image-based score following.
\end{itemize}

%% file: sections/related_work.tex
\section{Related Work}\label{sec:related}

\subsection{Symbolic Score Following}
Classical real-time score following relies on symbolic scores such as MIDI or MusicXML. Online DTW \cite{dixon2005line} provides a straightforward baseline by incrementally matching audio frames to score events. Probabilistic state-space models offer richer temporal modeling: Cont \cite{cont2010coupled} introduced a coupled duration-focused architecture, while Arzt and Widmer \cite{arzt2010towards} addressed robustness under arbitrary performer deviations using multi-agent tracking. Particle-filter methods such as that of Korzeniowski et al.\ \cite{korzeniowski2013tracking} extended the probabilistic framework to handle tempo changes and rests. Jiang and Raphael \cite{jiang2020score} proposed a switching state-space model that tracks hidden tempo changes, improving alignment stability under expressive performance variation. Nakamura et al.\ \cite{nakamura2016real} modeled repeats and skips with an explicit transition structure. Their analysis further revealed that the majority of real-world score jumps are preceded by short silent breaks, an observation that motivates the jump recovery design in Section~\ref{subsec:jump}. More recently, Peter et al.\ \cite{peter2025pairing} combined real-time piano transcription with symbol-level tracking, achieving strong results by leveraging automatic music transcription as a front end. Matchmaker \cite{park2025matchmaker} provides a systematic evaluation framework for symbolic methods, though it does not cover image-based trackers.

\subsection{Image-Based Score Following}
Dorfer et al.\ \cite{dorfer2016towards} first showed that a multimodal convolutional neural network (CNN) can follow sheet music without symbolic input, classifying position into discrete staff-level buckets on small image crops. Reinforcement-learning formulations then framed the task as agent navigation over score strips \cite{dorfer2018rl,henkel2019score}. Henkel et al.\ \cite{henkel2020learning} moved to full-page processing via audio-conditioned U-Net segmentation, predicting a pixel mask whose center of mass gives the note position. CYOLO-SB \cite{henkel2021real} extended this to multi-resolution detection: three independent YOLO-style heads predict note, bar, and system bounding boxes from feature maps conditioned via Feature-wise Linear Modulation (FiLM)~\cite{perez2018film}. However, the three heads predict in parallel without cross-level constraints, predictions at each frame are decoded without temporal consistency constraints across frames, and no recovery mechanism exists for when tracking is lost.
\begin{figure*}[t]
\centering
\includegraphics[width=\textwidth]{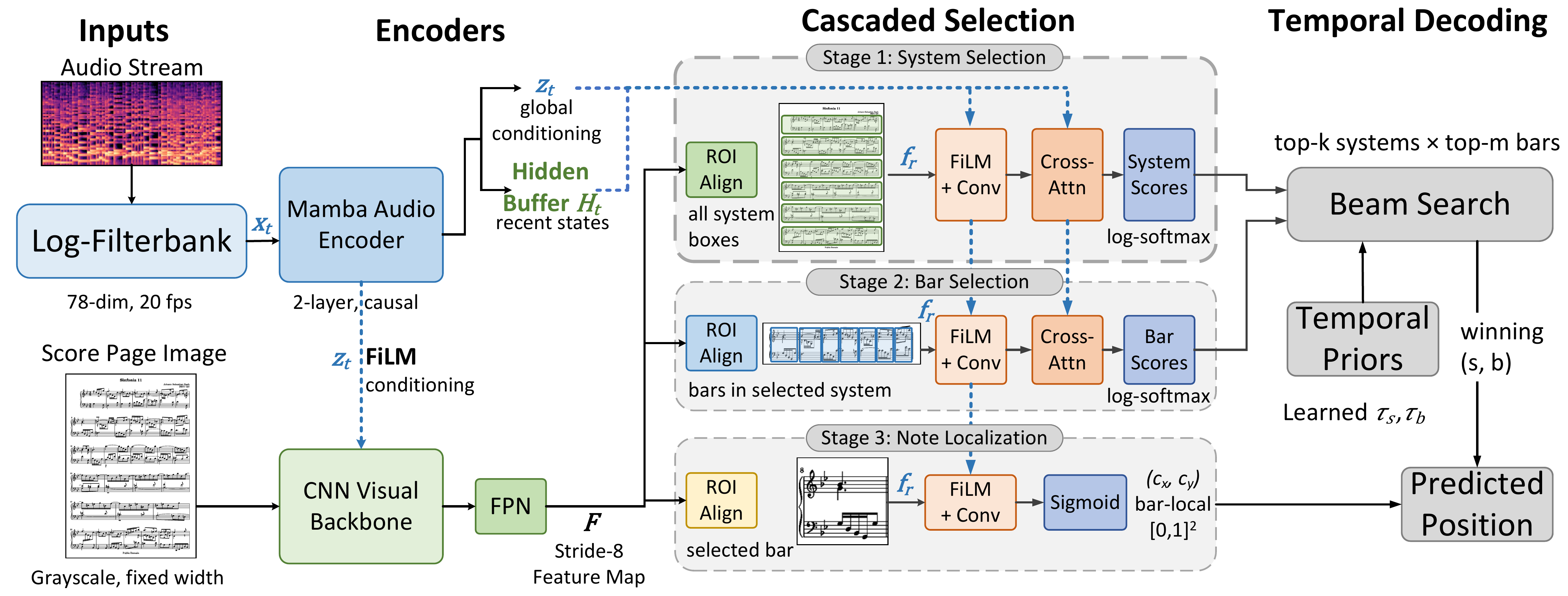}
\caption{Overview of the CODA architecture. A causal Mamba audio encoder and a shared CNN visual backbone feed into three cascaded stages: system selection, bar selection, and note localization. Beam search with learned temporal priors decodes the cascade over time.}
\label{fig:overview}
\end{figure*}

\subsection{Handling Repeats and Jumps}
Score discontinuities, including repeats, D.C., and coda jumps, pose a long-standing challenge for score following systems. In the symbolic domain, Nakamura et al.\ \cite{nakamura2016real} handle arbitrary repeats and skips through explicit state transitions. For offline audio-to-sheet-image synchronization, Shan and Tsai \cite{shan2020improved} proposed hierarchical DTW over learned audio and image features to handle repeats and jumps without requiring prior knowledge of jump locations. Morsi and Serra \cite{morsi2022bottlenecks} identified repeat handling as one of the key bottlenecks in audio-to-score alignment research. However, no existing real-time image-based score following method includes an explicit mechanism for recovering from score discontinuities during online tracking.

%% file: sections/method.tex
\section{Method}\label{sec:method}

\subsection{Problem Formulation}

CODA operates on the entire score page. Layout metadata provides a finite set of systems and bars, along with their bounding boxes, for each page. The model selects among candidate systems and bars on the current page.

At each audio frame $t$, let $h_t = (x_{\le t},\, I)$ denote the causal audio history and the score image for the current page. The model predicts three quantities: the active system index $s_t$, the active bar index $b_t$ within that system, and a continuous note position $u_t = (c_x, c_y)$ expressed in bar-local coordinates. Rather than predicting all three independently, CODA factorizes the joint distribution as
\begin{equation}
\begin{split}
p(s_t, b_t, u_t \mid h_t) = {} & p(s_t \mid h_t) \cdot p(b_t \mid s_t, h_t) \\
& \cdot\; p(u_t \mid b_t, s_t, h_t).
\end{split}
\end{equation}
This cascaded factorization enforces valid geometric structure by construction: the predicted bar always lies within the predicted system, and the predicted note always lies within the predicted bar.

\subsection{Model Architecture}

Figure~\ref{fig:overview} illustrates CODA's overall architecture. CODA has two input streams: audio and visual. The audio stream is converted into a 78-dimensional log-filterbank at 20~frames per second. A two-layer causal Mamba encoder~\cite{gu2024mamba} processes each frame. Its recurrent state compactly accumulates the entire audio history up to frame~$t$, yielding a conditioning vector~$z_t$. In parallel, a sliding window $H_t$ of the most recent $L$ per-frame outputs is maintained. $z_t$ provides conditioning for the visual feature map via FiLM, while $H_t$ serves as the key and value source for cross-attention between candidate visual regions and the recent audio history (Section~\ref{sec:cascade}).
 
The visual stream takes the score page image as input. The score page image is processed by a CNN backbone with FiLM conditioning driven by~$z_t$ at deeper stages. The output of the CNN backbone is further processed by a Feature Pyramid Network (FPN) that produces a stride-8 feature map~$F$ (i.e., at one-eighth the spatial resolution of the input). This feature map is shared across all three cascade stages (Section~\ref{sec:cascade}).

\subsection{Cascaded Selection}
\label{sec:cascade}
\begin{figure}[h]
\centering
\includegraphics[width=\columnwidth]{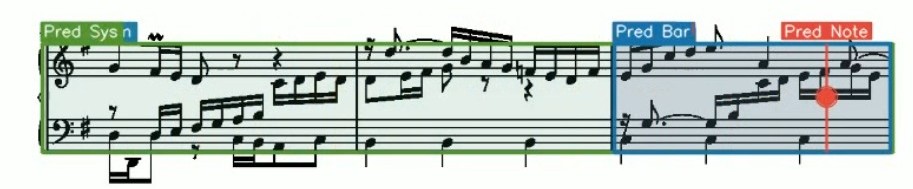}
\caption{CODA tracking output on an MSMD score page.}
\label{fig:tracking}
\end{figure}

Given the shared stride-8 feature map~$F$ from the visual input stream and the Mamba encoder outputs from the audio input stream, CODA applies three cascaded stages to localize the active position on the score page. Each stage narrows the spatial scope before proceeding to the next stage. All three stages share a common processing pipeline, with stage-specific variations detailed below. Figure~\ref{fig:tracking} illustrates the output on a real score page: the selected system (green), bar (blue), and note position (red) are each geometrically contained within the previous level.

Recall that $H_t \in \mathbb{R}^{L \times d}$ is the sliding window of the most recent $L$ Mamba outputs, with $d$ as the hidden dimension of the encoder. For each candidate region~$r$ (a system or bar bounding box from the score layout metadata), the pipeline proceeds as follows.

First, Region of Interest (ROI) Align \cite{he2017mask} extracts a fixed-size feature patch $f_r = \mathrm{ROIAlign}(F, r)$ from the stride-8 feature map. The extracted features are modulated by the audio vector~$z_t$ via FiLM as $\tilde{f}_r = \gamma(z_t) \odot f_r + \beta(z_t)$, where $\gamma$ and $\beta$ are learned linear projections and $\odot$ denotes element-wise multiplication. Convolutional layers further process the conditioned features: $\hat{f}_r = \mathrm{Conv}(\tilde{f}_r)$. In the system and bar stages, the convolved features are refined by scaled dot-product cross-attention \cite{vaswani2017attention} over the audio buffer: $\bar{f}_r = \mathrm{softmax}\!\left(\mathrm{flat}(\hat{f}_r)\,(W_a H_t)^\top / \sqrt{C}\right) W_a H_t$, where queries are formed by spatially flattening $\hat{f}_r$, keys and values are shared projections of $H_t$, and $W_a$ is a learned linear projection onto the $C$-dimensional attention space shared with $\hat{f}_r$. This cross-attention lets each candidate region attend to fine-grained temporal patterns in the recent audio history.

The system stage applies the full pipeline above (ROI Align, FiLM, convolution, cross-attention) to every system candidate on the page, producing~$\log p(s_t \mid h_t)$ via adaptive average pooling, a linear classifier, and softmax normalization. The bar stage applies the same pipeline with independent parameters to the bars within the selected system, yielding~$\log p(b_t \mid s_t, h_t)$. The note stage applies only ROI Align, FiLM, and convolution (no cross-attention at this stage) on the selected bar and regresses bar-local coordinates~$u_t = (c_x, c_y) \in [0,1]^2$ via sigmoid activation.

\subsection{Beam Search and Temporal Priors}

At each audio frame, the cascaded selection described above produces a log-probability over systems and another log-probability over bars within each system. To decode these scores over time, CODA uses beam search combined with learnable temporal priors.

At each frame~$t$, the model first scores every system on the page and retains the top-$k$ candidates. For each candidate system, the bars within that system are scored, and the top-$m$ candidates are retained per system. The actual values of $k$ and $m$ are specified in Section~\ref{sec:setup}. The note head is evaluated only on the winning system--bar pair after the beam has been resolved. This two-level pruning keeps inference efficient: the full system distribution is computed once per frame, but bar-level and note-level computation is restricted to the $k \times m$ beam.

The beam ranking combines model confidence with learned temporal priors that encode transition preferences between consecutive frames. Specifically, the composite score for a system--bar hypothesis~$(s, b)$ at frame~$t$ is
\begin{equation}
\begin{split}
\mathrm{score}_t(s,b) = {} & \log p(s_t \mid h_t) + \tau_s(s_t, s_{t-1}) \\
& + \log p(b_t \mid s_t, h_t) + \tau_b(b_t, b_{t-1}),
\end{split}
\end{equation}
where $\tau_s$ and $\tau_b$ are page-local transition penalties for systems and bars, respectively. These priors are implemented as learnable parameters that are trained end-to-end with the rest of the model. They are clamped to remain non-positive, with the stay transition (i.e., $s_t = s_{t-1}$ or $b_t = b_{t-1}$) fixed at zero. This parameterization biases the tracker toward smooth temporal progression by encouraging smooth transitions and penalizing large jumps. However, the penalties are bounded rather than hard-coded, so the model can still override them when the audio strongly indicates that a real jump is occurring. The jump detection and recovery mechanism will be covered in Section~\ref{subsec:jump}.

The hypothesis with the highest composite score determines the predicted system~$\hat{s}_t$ and bar~$\hat{b}_t$. The note head then regresses the note position~$\hat{u}_t$ within the selected bar, and the final output is the triplet~$(\hat{s}_t, \hat{b}_t, \hat{u}_t)$.

\subsection{Training}

Because CODA is cascaded, the bar and note stages only consider candidates within the selected system. This cascaded dependency means that if the system prediction is wrong, all downstream predictions will also be wrong. If the model is trained exclusively with ground-truth system labels, it never learns to handle incorrect system inputs, creating a mismatch between training and inference.

To address this, training proceeds in two phases using scheduled sampling~\cite{bengio2015scheduled}. In the first phase, the bar and note stages always receive the ground-truth system labels from training data. This allows the bar head to focus on discriminating among bar candidates under ideal system selections. In the second phase, starting from the first-phase checkpoint, the model's own system prediction is gradually mixed in. At each training step in the second phase, the ground-truth system is replaced by the model's self-predicted system with probability~$p_{\mathrm{pred}}$, which increases linearly from~0 to a maximum value~$p_{\max}$ over training. When the predicted system is wrong and does not contain the ground-truth bar, the bar and note training losses for that frame are masked because there is no meaningful supervision target. The system loss remains active, so the system head continues to receive corrective gradients. This two-phase curriculum training schedule progressively closes the gap between training and inference conditions.

The system and bar stages are trained with cross-entropy loss over their respective candidate sets, and the note head is trained with mean squared error on the bar-local coordinates. The three task losses are combined via learned uncertainty weighting \cite{kendall2018multi}:
\begin{equation}\label{eq:loss}
\mathcal{L} = \frac{1}{2\sigma_s^2}\mathcal{L}_{\mathrm{sys}} + \frac{1}{2\sigma_b^2}\mathcal{L}_{\mathrm{bar}} + \frac{1}{2\sigma_n^2}\mathcal{L}_{\mathrm{note}} + \log \sigma_s \sigma_b \sigma_n,
\end{equation}
where $\sigma_s$, $\sigma_b$, and $\sigma_n$ are learnable task-specific uncertainty parameters that balance the three objectives automatically.

\subsection{Jump Detection and Recovery}
\label{subsec:jump}

\begin{figure}[h]
\centering
\includegraphics[width=\columnwidth]{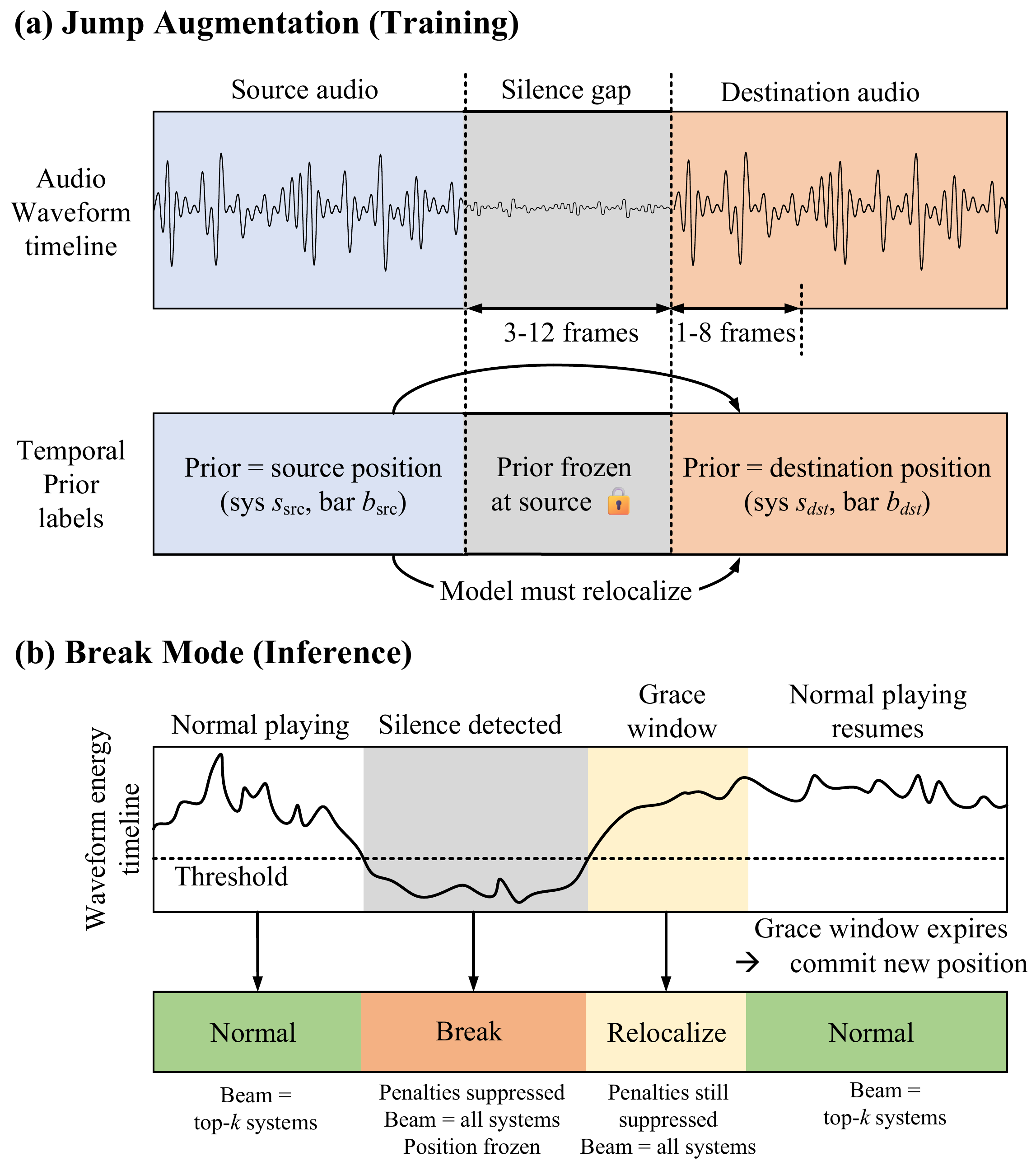}
\caption{Jump recovery mechanism. (a)~Jump augmentation during training. (b)~Break mode at inference.}
\label{fig:jump_recovery}
\end{figure}

Score discontinuities such as repeats, D.C., coda jumps, and unscripted performer errors move the active position non-monotonically. Rather than modeling each type, CODA uses silence as a cue to relax temporal constraints. Nakamura et al.~\cite{nakamura2016real} observed breaks longer than 500\,ms in 59 of 63 repeats/skips in clarinet practice recordings. Silence is therefore a practical recovery cue, not a universal requirement; Figure~\ref{fig:jump_recovery} shows the two mechanisms.

The jump-augmented training is illustrated in Figure~\ref{fig:jump_recovery}(a). Jump augmentation splices each sample with a controllable probability: a short silence (3--12 frames) is inserted, followed by audio from a randomly selected destination on the same or a different page. For same-page jumps, the previous system and bar labels are frozen at the source position, forcing the model to relocalize from audio evidence against a biased temporal prior.
 
At inference time, the break mode monitors waveform energy using a hysteresis rule. As presented in Figure~\ref{fig:jump_recovery}(b), when energy stays below a threshold for several consecutive frames, the tracker enters a break state: transition penalties are suppressed, the system beam is widened to cover all candidates, and the committed position is frozen. When energy rises again, penalties remain suppressed for a short grace window during which the tracker relocalizes to the jump destination. The Mamba hidden state is preserved throughout, maintaining temporal continuity across the silence gap.

%% file: sections/experiments_and_results.tex
\section{Experiments and Results}\label{sec:eval}

\subsection{Experimental Setup}\label{sec:setup}
 
\subsubsection{Dataset.}
Training and evaluation are both based on the MSMD dataset~\cite{dorfer2018learning}, using the preprocessed version provided by Henkel and Widmer~\cite{henkel2021real}, in which each piece is stored as a score image with per-note coordinate annotations and per-system and per-bar bounding box annotations, paired with a synthesized audio file (22,050~Hz). The standard split contains 354 training, 19 validation, and 94 test pieces.
 
\subsubsection{Evaluation protocol.}
We consider two evaluation settings matching~\cite{henkel2021real}: 
Setting~I uses the full 94-piece test split with synthetic score images 
and synthetic audio; Setting~II pairs the same synthetic images with real 
piano recordings for a 16-piece subset~\cite{henkel2019score}, testing 
generalization to real audio. Settings~III and~IV of~\cite{henkel2021real}, 
which require commercially published scanned scores, are not publicly 
available and are therefore not included.

Following~\cite{henkel2021real}, we evaluate at each ground-truth note onset.
Because image-based trackers predict positions in pixel space, we first convert each predicted position to the time domain by interpolating through the ground-truth onset-to-pixel mapping, then compute the absolute time difference between the predicted and true onset. We report the cumulative ratio of onsets tracked below five error thresholds: $\leq$\,0.05, 0.10, 0.50, 1.00, and 5.00~seconds. Higher ratios indicate better tracking. For methods that produce multi-resolution outputs, we additionally report \emph{system accuracy} and \emph{bar accuracy}, defined as the fraction of evaluated frames in which the predicted system or bar matches the ground-truth.
 
\subsubsection{Baselines.}
We compare against the image-based score following methods evaluated
in~\cite{henkel2021real}: \textbf{MM-Loc}~\cite{dorfer2016towards}, a
supervised multimodal localization model; \textbf{RL}~\cite{henkel2019score},
a reinforcement-learning agent; \textbf{CUNet}~\cite{henkel2020learning}, a
conditional U-Net for full-page segmentation; \textbf{CYOLO}~\cite{henkel2021multi}, the conditional
YOLO detector predicting note-level bounding boxes; and \textbf{CYOLO-SB}~\cite{henkel2021real}, the multi-resolution variant that adds system and bar heads. All baseline numbers are taken directly from~\cite{henkel2021real} under the same evaluation protocol. We also list \textbf{CYOLO-SB\,+\,A} for reference, noting that this variant uses additional proprietary training data (scanned scores and commercial recordings) that are not available to other methods. Methods that rely on automatic music transcription as a front end, such as Peter et al.~\cite{peter2025pairing}, are not directly comparable because they depend on symbolic-level intermediate representations rather than operating on sheet images.
 
\subsubsection{Implementation details.}
Audio is processed at 22,050~Hz with a 2048-sample short-time Fourier transform (STFT) (hop size 1\,102,
${\approx}\,$20~frames per second) and a 78-bin log-filterbank spanning
60\,Hz--6\,kHz. Score pages are converted to grayscale and scaled to a fixed
width of 416 pixels. The Mamba audio encoder has two layers with hidden
dimension~64, state dimension~16, convolution width~4, expansion factor~2, and projects to a 128-dimensional conditioning vector~$z_t$. Both the system and bar selection heads use cross-attention with four heads over an audio buffer of 64 frames. The beam search retains $k{=}3$ system and $m{=}3$ bar candidates per frame. Break mode applies normalized-energy onset and release thresholds of~0.1 and~0.25. The gap provides hysteresis against rapid toggling. Requiring 3 low-energy frames (${\approx}\,$150\,ms) filters transient dips, while an 8-frame (${\approx}\,$400\,ms) grace window provides post-resumption evidence. During break, priors are zeroed and the beam covers all systems, with $m{=}3$ bars per system.
 
Training follows the two-phase curriculum described in
Section~\ref{sec:method}. Phase~1 trains for 30~epochs with a learning rate of
$5{\times}10^{-4}$ under ground-truth system routing. Phase~2 fine-tunes for
20~epochs at $1{\times}10^{-4}$ with scheduled sampling, where the probability of using the model's own system prediction ramps linearly to $p_{\max}{=}0.7$ over the first 5~epochs. Both phases use the AdamW optimizer with a batch size of~16, cosine learning-rate decay, and gradient clipping at
norm~1.0. Data augmentation includes jump augmentation (Section~\ref{subsec:jump}), spatial shifts, tempo scaling, and cold-start truncation. Following~\cite{henkel2021real}, impulse response (IR) augmentation is also applied during training, convolving the audio with a randomly selected room impulse response on-the-fly to model varying microphone and room conditions. Jump augmentation samples destinations from a weighted mixture of six categories: \emph{repeat} (40\%), \emph{bar correction} (15\%), \emph{skip} (15\%), \emph{restart} (10\%), \emph{page jump} (10\%), and \emph{random} (10\%).
 
CODA has 2.0M trainable parameters. All training and experiments are conducted on a single NVIDIA RTX A6000 GPU (48\,GB) with an Intel Xeon Gold 5218R CPU and 64\,GB RAM. At inference, CODA processes each audio frame in 12.8\,ms (78.1\,fps), within the 50\,ms real-time budget at 20\,fps.

\subsection{Standard Tracking Results}\label{sec:tracking}
 
Table~\ref{tab:main} compares CODA to the baselines on the two evaluation
settings described in Section~\ref{sec:setup}. In Setting~I, CODA achieves .914 at ${\leq}\,0.10$\,s compared to .837 for CYOLO-SB, with bar accuracy improving from .890 to .975 and system accuracy from .963 to .991. In Setting~II, CODA achieves .743 at ${\leq}\,0.10$\,s versus .630 for CYOLO-SB.

\begin{table}[h]
\centering
\scriptsize
\setlength{\tabcolsep}{2.5pt}
\begin{tabular}{@{}l ccccc cc@{}}
\toprule
& \multicolumn{5}{c}{Onset error threshold (s)} & \multicolumn{2}{c}{Frame accuracy} \\
\cmidrule(lr){2-6} \cmidrule(l){7-8}
Method & ${\leq}$.05 & ${\leq}$.10 & ${\leq}$.50 & ${\leq}$1.0 & ${\leq}$5.0 & Bar & Sys \\
\midrule
\rowcolor{sechl}
\multicolumn{8}{@{}l}{\textit{Setting I\!: Synthetic images -- Synthetic audio}} \\
MM-Loc~\cite{dorfer2016towards}       & .707 & .747 & .839 & .855 & .917 & --   & --   \\
RL~\cite{henkel2019score}             & .411 & .435 & .776 & .856 & .971 & --   & --   \\
CUNet~\cite{henkel2020learning}       & .726 & .750 & .855 & .885 & .937 & --   & --   \\
CYOLO~\cite{henkel2021multi}          & .830 & .842 & .885 & .909 & \second{.984} & --   & --   \\
CYOLO-SB~\cite{henkel2021real}        & .820 & .837 & .893 & .912 & .983 & .890 & \second{.963} \\
CYOLO-SB+A$^\dagger$~\cite{henkel2021real} & \second{.846} & \second{.861} & \second{.908} & \second{.927} & \second{.984} & \second{.892} & .956 \\
\rowcolor{rowhl}
CODA (ours)  & \first{.897} & \first{.914} & \first{.955} & \first{.981} & \first{.996} & \first{.975} & \first{.991} \\
\midrule
\rowcolor{sechl}
\multicolumn{8}{@{}l}{\textit{Setting II\!: Synthetic images -- Real audio recordings}} \\
MM-Loc~\cite{dorfer2016towards}       & .364 & .406 & .585 & .611 & .735 & --   & --   \\
RL~\cite{henkel2019score}             & .185 & .200 & .476 & .603 & .901 & --   & --   \\
CUNet~\cite{henkel2020learning}       & .113 & .125 & .224 & .266 & .443 & --   & --   \\
CYOLO~\cite{henkel2021multi}          & .563 & .581 & .712 & .749 & .919 & --   & --   \\
CYOLO-SB~\cite{henkel2021real}        & .610 & .630 & .799 & .832 & .960 & .829 & .917 \\
CYOLO-SB+A$^\dagger$~\cite{henkel2021real} & \second{.682} & \second{.706} & \second{.865} & \second{.891} & \second{.981} & \second{.865} & \second{.941} \\
\rowcolor{rowhl}
CODA (ours)  & \first{.721} & \first{.743} & \first{.883} & \first{.914} & \first{.987} & \first{.891} & \first{.953} \\
\bottomrule
\end{tabular}
\caption{Score-following comparison. Each cell is the ratio of onsets tracked below the error threshold (higher is better); best in \first{red bold}, second best in \second{blue underline}. $^\dagger$Proprietary training data; baseline numbers from~\cite{henkel2021real}.}
\label{tab:main}
\end{table}

\vspace{-0.7cm}

\subsection{Jump Recovery Evaluation}\label{sec:jump}
 
To fairly evaluate jump recovery, we construct a jump test benchmark from the 94 MSMD test pieces. We manually annotated the repeat structure of each piece (repeat barlines, da capo, volta brackets, binary form, etc.). Of the 94 pieces, 66 contain written repeat structures and 28 do not. The benchmark is partitioned into two subsets accordingly. The \emph{repeat subset} comprises the 66 pieces with repeats, producing 131 jumps that follow the annotated performance order. The \emph{random subset} comprises the remaining 28 pieces without repeats, each receiving 3 randomly placed jumps (84 jumps in total) as a stress test.
 
We report three metrics on post-jump segments: \emph{system recovery rate}
(fraction of jumps where the correct system is identified within 1.0 and
2.0\,s), \emph{mean recovery latency} (time in seconds from audio resumption to the first correct system
prediction), and \emph{post-jump tracking accuracy} (${\leq}\,1.0$\,s threshold
in the 5\,s window after each jump).
 
\begin{table}[h]
\centering
\scriptsize
\setlength{\tabcolsep}{2.5pt}
\begin{tabular}{@{}ll cc c c@{}}
\toprule
& & \multicolumn{2}{c}{Sys recovery} & & Post-jump \\
\cmidrule(lr){3-4}
Subset & Method & @1\,s & @2\,s & Lat.\,(s) & ${\leq}$1\,s \\
\midrule
\multirow{3}{*}{Repeat}
 & CYOLO-SB       & .12   & .20   & 4.31   & .35 \\
 & CODA w/o break & \second{.29}   & \second{.44}   & \second{2.63}   & \second{.54} \\
\rowcolor{rowhl}
 & CODA (full)    & \first{.78} & \first{.91} & \first{0.72} & \first{.82} \\
\midrule
\multirow{3}{*}{Random}
 & CYOLO-SB       & .08   & .15   & 5.18   & .27 \\
 & CODA w/o break & \second{.23}   & \second{.38}   & \second{3.07}   & \second{.46} \\
\rowcolor{rowhl}
 & CODA (full)    & \first{.64} & \first{.80} & \first{1.24} & \first{.71} \\
\bottomrule
\end{tabular}
\caption{Jump recovery results on the repeat-aware MSMD test set.
\textbf{Rec.@$k$\,s}: fraction of jumps with correct system within $k$\,s.
\textbf{Lat.}: mean seconds to first correct system prediction.
\textbf{Post-J.}: fraction of onsets tracked within ${\leq}\,1.0$\,s in the 5\,s window after each jump.}
\label{tab:jump}
\end{table}
 
Table~\ref{tab:jump} shows the results. CYOLO-SB yields low recovery across both subsets, while CODA without break mode struggles because temporal penalties resist large jumps. Full CODA reaches .78 at 1\,s on repeat versus .64 on random. Annotated repeats resume at musically defined bar boundaries, while random jumps may resume at arbitrary onsets and therefore form a harder stress test.

\subsection{Ablation Study}\label{sec:ablation}

To isolate the contribution of each component, we evaluate five ablated variants of CODA on Setting~I (synthetic images and audio). Each variant removes or disables a single component while keeping all others unchanged.

Table~\ref{tab:ablation} shows the results. The largest degradation comes from removing the cascade ($-.045$ at ${\leq}\,0.10$\,s, bar accuracy drops from .975 to .931). Beam search, cross-attention, and temporal priors contribute the next-largest gains ($-.026$, $-.019$, and $-.014$ at ${\leq}\,0.10$\,s, respectively). Scheduled sampling has the smallest individual effect, but consistently improves all thresholds.

\begin{table}[h]
\centering
\scriptsize
\setlength{\tabcolsep}{2.5pt}
\begin{tabular}{@{}l cccc cc@{}}
\toprule
& \multicolumn{4}{c}{Onset error threshold (s)} & \multicolumn{2}{c}{Frame accuracy} \\
\cmidrule(lr){2-5} \cmidrule(l){6-7}
Variant & ${\leq}$.05 & ${\leq}$.10 & ${\leq}$.50 & ${\leq}$1.0 & Bar & Sys \\
\midrule
\rowcolor{rowhl}
CODA (full)            & \first{.897} & \first{.914} & \first{.955} & \first{.981} & \first{.975} & \first{.991} \\
\midrule
w/o cascade            & .851 & .869 & .923 & .952 & .931 & .967 \\
w/o cross-attention    & .878 & .895 & .943 & .972 & .958 & .985 \\
w/o beam search        & .870 & .888 & .938 & .968 & .949 & .982 \\
w/o temporal priors    & .882 & .900 & .948 & .975 & .963 & .988 \\
w/o sched.\ sampling   & \second{.888} & \second{.906} & \second{.951} & \second{.978} & \second{.968} & \second{.989} \\
\bottomrule
\end{tabular}
\caption{Ablation study on the MSMD synthetic test set (Setting~I). Each row
removes one component from the full model.}
\label{tab:ablation}
\end{table}

%% file: sections/conclusion.tex
\section{Discussion and Conclusion}\label{sec:conclusion}
This paper presents CODA, which formulates image-based score following as
cascaded selection over known candidates and introduces a silence-driven
break mode for jump recovery. We also contribute a repeat-aware jump test benchmark with manually annotated repeat structures for the MSMD test set, providing the first standardized evaluation protocol for discontinuity handling in image-based score following. Despite achieving state-of-the-art performance, several limitations remain. CODA is trained and evaluated exclusively on solo piano from the MSMD dataset~\cite{dorfer2018learning}. Larger datasets that pair real recordings with scanned scores would benefit not only this work but the field as a whole. Extending CODA to other solo instruments is a natural first step. Beyond solo, chamber music and orchestral settings pose additional challenges due to timbral overlap, denser layouts, and multiple simultaneous parts. Finally, evaluation under real-world conditions, including scanned pages, commercial recordings, and live microphone input, is needed to further assess the method.

%% file: sections/acknowledgments.tex
\section{Acknowledgments}
This work was supported by the Natural Sciences and Engineering Research Council of Canada (NSERC; project numbers RES0048688, RES0051374, and RES0054326) and Alberta Innovates (project number RES0053965). Computational resources were provided by the research group of Jie Han at the University of Alberta.